\documentclass[nofootinbib,aps,preprint,superscriptaddress]{revtex4}
\usepackage{amsmath}
\usepackage{citesort}
\usepackage{epsfig}

\newcommand{\beqa}{\begin{eqnarray}}
\newcommand{\eeqa}{\end{eqnarray}}
\newcommand{\beq}{\begin{equation}}
\newcommand{\eeq}{\end{equation}}
\newcommand{\bal}{\begin{align}}
\newcommand{\eal}{\end{align}}

\def\gsim{\ \rlap{\raise 3pt \hbox{$>$}}{\lower 3pt \hbox{$\sim$}}\ }
\def\lsim{\ \rlap{\raise 3pt \hbox{$<$}}{\lower 3pt \hbox{$\sim$}}\ }

\def\DbarZ{\bar{D}^0}

\def\Pex{ P_{\rm ex} }
\def\Psign{P_-}
\def\Ppi{P_\pi}

\begin{document}

\preprint{\vbox{
\hbox{}
\hbox{SLAC-PUB-10323}
\hbox{TECHNION-PH-2004-5}
\hbox{hep-ph/0402055}
\hbox{February 2004}
}}

\vspace*{48pt}

\title{Using untagged $B^0 \to D K_S$ to determine $\gamma$}

\def\addtech{Department of Physics,
Technion--Israel Institute of Technology,\\
Technion City, 32000 Haifa, Israel\vspace*{6pt}}
\def\addslac{Stanford Linear Accelerator Center, %\\
  Stanford University, Stanford, CA 94309\vspace*{6pt}}
\def\adducsc{Santa Cruz Institute for Particle Physics, %\\
  University of California, Santa Cruz, CA 95064\vspace*{6pt}}

\author{M. Gronau}\affiliation{\addtech}
\author{Y. Grossman}\affiliation{\addtech} \affiliation{\addslac}
\affiliation{\adducsc}
\author{N. Shuhmaher} \affiliation{\addtech}
\author{A. Soffer}
\affiliation{Department of Physics, Colorado State University,
Fort Collins, CO 80523\vspace*{6pt}}
\author{J. Zupan}
\affiliation{\addtech}
\affiliation{J.~Stefan Institute, Jamova 39, P.O. Box 3000,1001
Ljubljana, Slovenia\vspace*{18pt}}

\begin{abstract} \vspace*{18pt}
It is shown that the weak phase $\gamma \equiv 
\arg(-V_{ud}V^*_{ub}V_{cb}V_{cd}^*)$  can be determined
using only untagged decays $B^0/\bar B^0 \to D K_S$.
In order to reduce the uncertainty in  $\gamma$, we suggest 
combining information from $B^{\pm}\to DK^{\pm}$ and
from untagged $B^0$ decays, where the $D$
meson is observed in common decay modes. 
Theoretical assumptions, which may further 
reduce the statistical error, are also discussed.
\end{abstract}

\maketitle

%%%%%%%%%%%%%%%%%%%%%%%%%%%%%%%%%
%%%%%%%%%%%%%   I   %%%%%%%%%%%%%
%%%%%%%%%%%%%%%%%%%%%%%%%%%%%%%%%
\section{Introduction}
CP violation measured in $B \to J/\psi K_S$~\cite{psiKS} is
interpreted in terms of the phase $\beta \equiv
\arg(-V_{tb}V^*_{td}V_{cd}V_{cb}^*)$ in a way which is practically free
of  theoretical uncertainties, providing an important test of the
Kobayashi-Maskawa mechanism~\cite{KM}.
On the other hand, the current interpretation of CP asymmetry 
measurements in $B \to \pi^+\pi^-$~\cite{Abe:2004us} in terms of the 
phase $\alpha \equiv \arg(-V_{td}V^*_{tb}V_{ub}V_{ud}^*)$ involves 
an uncertainty in the ratio of penguin-to-tree amplitudes~\cite{MGJR}. 
A theoretically clean method  \cite{GW,GL} for measuring the phase 
$\gamma \equiv \arg(-V_{ud}V^*_{ub}V_{cb}V_{cd}^*)$ 
involves interference between tree amplitudes
$\bar b \to \bar c u \bar s$ and $\bar b \to \bar u c \bar s$, governing
$B \to \DbarZ X_s$ and $B \to D^0 X_s$, where the $\DbarZ$ and $D^0$
decay to a common hadronic state, and $X_s = K, K^*, K\pi, \dots $ is a 
strangeness one state.

Originally, this idea for measuring $\gamma$ was proposed for 
charged $B$ decays~\cite{GW} and for time-dependent neutral $B$
decays~\cite{GL} of the type $B\to D_{\rm CP}K$.  Here positive
(negative) CP eigenstates, such as $K^+K^-$ ($K_S\pi^0$), identify
equal admixtures of $D^0$ and $\bar D^0$ states with equal
(opposite) signs.  A variety of other $D$ decay final
states can be used as well, leading to several variants of the original 
method~\cite{ADS,GLS}. Every hadronic 
state accessible at tree
level to $D^0$ decay is also accessible at tree level to $\DbarZ$
decay with varying levels of Cabibbo suppression.  Flavor states,
$K^-\pi^+$ and $K^{*-}\pi^+$, which are Cabibbo-favored in $D^0$ decays,
are doubly Cabibbo-suppressed in $\bar D^0$ decays.  Flavorless states,
such as $K^{*+}K^-$ and $K^+ K^{*-}$, are produced in singly
Cabibbo-suppressed decays of both $D^0$ and $\bar D^0$.  
Recently, it was shown that a model-independent
extraction of $\gamma$ is also possible by considering $B^{\pm} \to
DK^{\pm}$ with subsequent multibody $D$ decay, such as 
$D \to K_S \pi^+\pi^-$~\cite{ggsz,Atwood:2003mj}. Other variants
make use of the decays $B^{\pm} \to D^*K^{\pm}, D^* \to D \pi^0$,
the self-tagged decay mode $B^0 \to DK^{*0}, K^{*0}\to K^+\pi^-$~\cite{ID}, multibody 
$B$ decays of the type $B \to D K\pi$~\cite{multi} and combinations of these 
processes~\cite{Atwood:2003jb}.

First results for $B^{\pm} \to D_{\rm CP} K^{\pm}$ were presented
recently by the Belle~\cite{DCPBelle} and BaBar~\cite{DCPBabar}
collaborations. These studies were based on several tens of events in
each experiment and demonstrate the potential of a larger data sample
in providing useful constraints on $\gamma$~\cite{MGFPCP}. The Belle
collaboration has also presented a preliminary analysis of about one
hundred events of the type $B^{\pm} \to DK^{\pm}, D \to K_S
\pi^+\pi^-$~\cite{BelleDalitz}, from which constraints on
$\gamma$ were obtained. The main difficulty of each of these methods is 
that each decay mode by itself has a very low rate. 
Reaching high sensitivity in near future measurements of $\gamma$ 
requires combining several relevant $B$ and $D$ decay modes.

Measurements relevant to studying $\gamma$ have so far focused
on charged $B$ decays, $B^\pm \to D K^\pm$ and 
$B^\pm \to D K^{*\pm}$~\cite{DK*}. At first glance, neutral $B$ decays seem to be less
promising for two reasons: they have much smaller rates and require
$B^0$ flavor tagging. Let us discuss these two points one at a time:
\begin{itemize}
\item The processes $B^0 \to D K^0$ and $B^0 \to D K^{*0}$ are expected
to be color-suppressed~\cite{ID}, 
implying that their rates are about an order of magnitude below the rates
of corresponding charged $B$ decays. However, the crucial factor
determining the sensitivity of a measurement of $\gamma$ is not the decay
rate itself. Rather, the sensitivity is governed by the magnitude of
the smaller of the two interfering amplitudes in a given
process. Since the smaller amplitudes in $B^+$ and $B^0$ decays are
both color-suppressed, they are expected to be of comparable
magnitudes. Therefore, this by itself is not a limiting factor for
neutral $B$ decays.
\item Only time-integrated rates have so far been measured in $B^0 \to D K^0$, 
combining rates for $B^0 \to \DbarZ K^0$ and $\bar B^0 \to \DbarZ \bar 
K^0$~\cite{BelleD0K0}.  
It was shown in~\cite{GL,KaLo,AtSo} that a determination of 
$\gamma$ is possible from time-dependent measurements, for which one 
must tag the flavor of the initial $B^0$. 
The effective flavor tagging efficiency at $B$ factories is about 30\%
(and much smaller at hadron machines), resulting in a doubling of the
statistical error relative to the perfect-tag case.  As we show below,
$\gamma$ can be determined using untagged data alone. This makes use
of events that cannot be tagged or even events that are mis-tagged,
regaining a significant part of the sensitivity lost due to the low
effective tagging efficiency.
\end{itemize}

In the present paper we investigate what can be learned from untagged
decays $B^0/\bar B^0 \to D K_S$, where the $D$ meson is observed in
several decay modes. A potential lower bound on $|\cos\gamma|$ from
untagged decays, in which $D$ mesons are observed in CP-eigenstates,
was noted by Fleischer~\cite{Fleischer}. We will go beyond this bound
by showing that  $\gamma$ can actually be completely determined
in the range $0 < \gamma < \pi$,
using only untagged decays. In practice, it is useful to combine information 
from untagged neutral $B$ decays with information from charged $B$ 
decays, since the observables related to $D$ decays are common to both 
cases. This provides an overconstrained information, permitting a 
more accurate determination of $\gamma$ than when using $B^+$ decays 
alone.

The plan of the paper is as follows. 
In Section II we introduce
parallel notations for charged and neutral $B \to D K$ decays,
discussing briefly relative magnitudes of decay amplitudes in these
processes. Section III studies two-body and quasi two-body $D$ decays,
distinguishing between several classes of decay modes. We
show that $\gamma$ can be determined from untagged neutral $B$ decays
alone and derive an explicit expression for~$\tan^2\gamma$~in terms of
measurable rates. Multibody $D$ decays in $B^0 \to DK_S$ are studied
in Section IV. In Section V we discuss a way of reducing the number of
hadronic parameters by assuming isospin symmetry and by neglecting an
annihilation contribution in $B \to DK$.  Finally, Section VI
concludes. We also add an appendix studying time-dependence in $B^0(t)
\to (K_S\pi^+\pi^-)_DK_S$.

%%%%%%%%%%%%%%%%%%%%%%%%%%%%%%%%
%%%%%%%%%%%%%  II  %%%%%%%%%%%%%
%%%%%%%%%%%%%%%%%%%%%%%%%%%%%%%%
\section{Amplitudes in $B\to DK$ decays}

We define decay amplitudes of $B \to DK$ for charged $B$ mesons,
\beqa\label{Ac}
A(B^+ \to \DbarZ K^+) & \equiv & A_c~,\nonumber \\ 
A(B^+ \to D^0 K^+) & \equiv & A_c r_c e^{i(\delta_c + \gamma)}~,
\eeqa
and for neutral $B$ mesons,
\beqa\label{An}
A(B^0 \to \DbarZ K_S) & \equiv & A_n~,\nonumber\\
A(B^0 \to D^0 K_S) & \equiv & A_n r_n e^{i(\delta_n + \gamma)}~.
\eeqa
By convention, $A_i \ge 0, r_i \ge 0$, and $0 \le \delta_i \le 2\pi~(i = c, n)$.
Amplitudes for the CP conjugated decays have the same expressions, but
the phase $\gamma$ occurs with an opposite sign.

Let us discuss briefly the relevant ratios of amplitudes.
The amplitude $A_c$ is a combination of color-allowed and
color-suppressed contributions, while the amplitude $A_n$ is purely
color-suppressed. The ratio $A_n/A_c$ may be estimated in two ways
leading to comparable values.
We mention in each case the required approximation:
\begin{itemize}
\item Measurements of 
$B^+\to \DbarZ K^+$~\cite{DCPBelle,DCPBabar,Bornheim:2003bv} 
and $B^0 \to \DbarZ K^0$~\cite{BelleD0K0} imply 
\beq\label{An/Ac1}
\frac{A_n}{A_c} \simeq \sqrt{\frac{\Gamma(B^0/\bar B^0 \to \DbarZ K_S)}
{\Gamma(B^+\to \DbarZ K^+)} } = 0.25 \pm 0.07~.
\eeq 
In addition to a term $A^2_n$, the untagged rate in the numerator includes  
also a smaller term $A^2_nr^2_n$
which we neglect.
\item Using flavor SU(3)~\cite{SU3}, one may  relate $A_n/A_c$ to a 
corresponding ratio
measured in $B \to \bar D \pi$~\cite{color},
\beq\label{An/Ac2}
\frac{A_n}{A_c} \simeq
\sqrt{\frac{\Gamma(B^0\to \DbarZ\pi^0)}{\Gamma(B^+ \to \DbarZ\pi^+)}} =
0.24 \pm 0.01~.
\eeq
This relation is affected by SU(3) breaking corrections, and by a small 
%MG changed
exchange amplitude in 
$B^0 \to \DbarZ\pi^0$~\cite{Chiang:2002tv,DsK}. SU(3) breaking 
corrections, which are common in $A(B^+\to \DbarZ K^+)/A(B^+ \to 
\DbarZ \pi^+)$ and $A(B^0\to \DbarZ K_S)/A(B^0 \to \DbarZ \pi^0)$, 
cancel in (\ref{An/Ac2}) and do not affect this estimate.
\end{itemize}

It is more difficult to obtain reliable estimates for $r_c$ and $r_n$. The 
two parameters are  expected to be smaller than one since they contain 
a CKM factor $|V_{ub}V_{cs}/V_{cb}V_{us}|\simeq 0.4$. The 
parameter $r_c$ involves also an unknown color-suppression factor 
in $\bar b \to \bar u c \bar s$, while $r_n$ involves the ratio of 
color-suppression factors in $\bar b \to \bar u c \bar s$ and  
$\bar b \to \bar c u\bar s$. 
Since the dynamics of $B \to \bar DK$ decays (caused by $\bar b \to \bar 
c u\bar s$) and $B \to DK$ (caused by $\bar b \to \bar u c \bar s$) are different, 
color-suppression may be different in the two cases. This introduces large 
uncertainties in $r_c$ and $r_n$
%MG added reference
\cite{BaBarADS}. 

It is easier to justify an approximate relation between the 
magnitudes of the two color-suppressed amplitudes 
$A_cr_c$ and $A_nr_n$. Noting that the two processes $B^+ \to D^0 K^+$ 
and $B^0 \to D^0 K^0$ differ only by the flavor of the spectator quark, one expects
\beq
A_c r_c \simeq \sqrt{2} A_n r_n~.
\eeq
In Section V we will discuss the approximation involved in this relation and 
a way of testing it experimentally. This approximate equality implies 
that the sensitivity to $\gamma$ is comparable in charged and neutral
$B$ decays, since the sensitivity in each case is governed by the 
smaller of the two interfering amplitudes. This point provides a major
motivation for our study. 

%%%%%%%%%%%%%%%%%%%%%%%%%%%%%%%%%
%%%%%%%%%%%%%   III %%%%%%%%%%%%%
%%%%%%%%%%%%%%%%%%%%%%%%%%%%%%%%%
%
\section{Two-body and quasi two-body $D$ decays}
Considering decays of $\DbarZ$ and $D^0$ into a generic two-body or
quasi two-body hadronic state $f_D$ and its CP conjugate $\bar f_D$, we denote
the corresponding amplitudes by
\beqa
A(\DbarZ \to f_D) & = & A(D^0 \to \bar f_D)  \equiv  A_f~,\nonumber\\
\label{Af}
A(D^0 \to f_D) & = & A( \DbarZ \to \bar f_D) \equiv A_f r_f e^{i\delta_f}~,
\eeqa
where by convention $A_f \ge 0$, $r_f \ge 0$, and $0 \le \delta_f \le
2\pi$. Here and below we set the weak phase in $D$ decays to zero and
neglect $D^0-\DbarZ$ mixing.  The effects of $D^0-\DbarZ$ mixing can
be included as in~\cite{DDbar}, 
but are not further discussed here.

Using these notations, one finds expressions for decay rates in charged
$B$ decays,
\beqa
\Gamma(B^+ \to f_D K^+) & = & A_c^2 A^2_f \left [ 1 + r_c^2 r^2_f
+ 2r_cr_f \cos(\delta_c + \delta_f + \gamma)\right ] ~,\nonumber\\
\Gamma(B^- \to \bar f_D K^-) & = & A_c^2 A^2_f \left [ 1 + r_c^2 r^2_f
+ 2r_c r_f \cos(\delta_c + \delta_f - \gamma)\right ] ~,\nonumber\\
\Gamma(B^+ \to \bar f_D K^+) & = & A_c^2 A^2_f \left [ r_c^2  + r^2_f
+ 2r_cr_f \cos(\delta_c - \delta_f + \gamma)\right ] ~,\nonumber\\
\label{GammaB-f}
\Gamma(B^- \to f_D K^-) & = & A_c^2 A^2_f \left [ r_c^2  + r^2_f
+ 2r_c r_f \cos(\delta_c - \delta_f - \gamma)\right ] ~.
\eeqa
Combining $B^+$ and $B^-$ decay rates for states involving a common
$D$ decay mode, $f_D$ or $\bar f_D$, one finds
\beqa
\langle \Gamma(B \to f_D K_c) \rangle & \equiv &
\Gamma(B^+ \to f_D K^+) + \Gamma(B^- \to f_D K^-)\nonumber\\
& = & A_c^2 A^2_f \left [ (1 + r_c^2)(1 + r_f^2) + 4 r _cr_f
\cos(\delta_f + \gamma)
\cos\delta_c \right ]~,\nonumber\\
\label{GammaB+-fbar}
\langle \Gamma(B \to \bar f_D K_c) \rangle & \equiv &
\Gamma(B^+ \to \bar f_D K^+) + \Gamma(B^- \to \bar f_D K^-)\nonumber\\
& = & A_c^2 A^2_f \left [ (1 + r_c^2)(1 + r_f^2) + 4 r_cr_f
\cos(\delta_f - \gamma)
\cos\delta_c \right ]~.
\eeqa

Studying neutral $B$ decays, one finds similar expressions for
untagged decay rates~\cite{KaLo}:
\beqa
\langle \Gamma(B \to f_D K_n) \rangle & \equiv &
\Gamma(B^0 \to f_D K_S) + \Gamma(\bar B^0 \to f_D K_S)\nonumber\\
& = & A^2_n A^2_f \left [ (1 + r^2_n)(1 + r_f^2) + 4 r_n r_f
\cos(\delta_f + \gamma)
\cos\delta_n \right ]~,\nonumber\\
\label{GammaBfbar}
\langle \Gamma(B \to \bar f_D K_n) \rangle & \equiv &
\Gamma(B^0 \to \bar f_D K_S) + \Gamma(\bar B^0 \to \bar f_D K_S)\nonumber\\
& = & A^2_n A^2_f \left [ (1 + r^2_n)(1 + r_f^2) + 4 r_n r_f
\cos(\delta_f - \gamma)
\cos\delta_n \right ]~.
\eeqa
Individual time-dependent decay rates for $B^0(t) \to f_D K_S,~\bar
B^0(t) \to f_D K_S$ and their CP-conjugates are given in~\cite{KaLo},
and include more information than the untagged rates.
These, however, will not be needed in the following.

The decay rates in Eqs.~(\ref{GammaB+-fbar}) and~(\ref{GammaBfbar})
display a dependence on two types of quantities. Amplitudes and strong
phases in $B \to DK$, ($A_i, r_i, \delta_i;~i=c,n$), which in general
obtain different values in charged and neutral $B$ decays, and the
corresponding quantities in $D^0/\DbarZ \to f_D$, ($A_f, r_f,
\delta_f$), which are common to both $B^+$ and $B^0$ decays.  We will refer
to these quantities as $B$ and $D$ decay parameters, respectively.  In
the following we will assume that the $D$ decay quantities $A_f$ and
$r_f$ have been measured and are known. They can be obtained through
branching ratio measurements in an independent sample of neutral $D$
mesons, flavor-tagged through their production in the decay $D^{*+}\to
D^0 \pi^+$~\cite{DCS+DDbar}. 
In quasi two-body $D$ decays the three parameters $A_f,r_f$ and 
$\delta_f$ can be determined simultaneously through a complete Dalitz 
plot analysis. Although the phases $\delta_f$ can in principle be
measured~\cite{Atwood:2003mj,deltaf,GGR,RS}, we will treat 
them as unknown, unless indicated otherwise. 

Note that the combined $B^{\pm}$ rates in~(\ref{GammaB+-fbar})
and those in the untagged $B^0$ decays~(\ref{GammaBfbar}) depend
in each case only on two combinations of $B$ decay parameters,
\beq\label{XY}
X_i  \equiv  A_i^2(1 + r_i^2)~,\qquad Y_i \equiv 2A_i^2 r_i\cos\delta_i~,
\qquad i = c, n~,
\eeq
which obey
\beq\label{Y<X}
|Y_i| \le X_i~.
\eeq
We see that the individual branching ratios for $B^0\to \DbarZ K_S$
and $B^0 \to D^0 K_S$, proportional to $A^2_n$ and $A^2_nr^2_n$,
respectively, cannot be measured from untagged decays alone.

The three $D$ decay parameters $A_f, r_f$ and $\delta_f$ in~(\ref{Af})
depend, of course, on the final state $f_D$. One may distinguish
between three cases for which we give examples:
\begin{enumerate}
\item $f_D =$ CP-state ({\it e.g.}, $f_{{\rm CP}+}=K^+K^-,~f_{{\rm CP}-}=
K_S\pi^0$), for which $r_{{\rm CP}\pm} =1, \cos\delta_{{\rm CP}\pm} = \pm 1$,
\item $f_D =$ flavorless ({\it e.g.}, $K^{*+}K^-$), for which
$r_f = {\cal O}(1)$ but generally $r_f \ne 1, \delta_f =$ unknown,
\item $f_D =$ flavor state ({\it e.g.}, $K^+\pi^-$), for which 
$r_f \simeq \tan^2\theta_C$~\cite{DCS+DDbar}, $\delta_f =$ unknown, where 
$\theta_C$ is the Cabibbo angle.
\end{enumerate}

Using Eqs. \eqref{GammaBfbar} and the observation \eqref{XY}, it is
simple to show that $\gamma$ may be determined solely from untagged
$B^0$ decays. Consider $N$ different non-CP neutral $D$ decay modes 
$f_D^k$ ($k = 1,...,N$)
together with their CP conjugates $\bar f_D^k$, as the final states 
in the $B^0 \to DK_S$ decay chain.  The unknown variables are
$\gamma,~X_n,~Y_n$ and $N$ strong phases
$\delta_f^k$. Eqs.~(\ref{GammaBfbar}), which provide $2N$ measurables
for $N + 3$ unknowns, are solvable for $N \ge 3$. That is, $\gamma$
may be determined from untagged $B^0\to DK_S$ decay rates, where $D^0$
is observed in at least three different non-CP decay modes and their
CP conjugates. This argument may be generalized to include other
untagged $B^0$ decays, such as $B^0 \to D^* K_S~(D^* \to
D^0\pi^0)$. Assuming $M$ different $B^0$ decay modes of this kind,
each of which introduces a pair of unknowns  $X_n^j$ and
$Y_n^j$ ($j = 1,...,M$), one has $2MN$ measurables for $2M + N +1$ 
unknowns. For $M \ge
2$ this set of equations is solvable for $N \ge 2$. Namely, two non-CP
decay modes of $D^0$ are sufficient for determining $\gamma$ from
untagged $B^0 \to D K_S$ and $B^0\to D^* K_S$.

For a CP-eigenstate the strong phase $\delta_{f({\rm CP})}$ is either $0$ or
$\pi$. In this case the two equations in~(\ref{GammaBfbar}) become 
identical and provide a single measurable. Choosing the decay modes $f_D^k$ 
to be (i) an even-CP state, (ii) an odd-CP state, and (iii) a single non-CP 
eigenstate and its CP conjugate (involving an unknown phase $\delta_f$), one 
can solve the four equations for $\gamma,~X_n,~Y_n$ and $\delta_f$.
For this case we now derive an explicit expression for $\tan^2\gamma$ in
terms of measurable rates. The derivation holds for both charged and 
neutral $B$ decays.

Using Eqs.~(\ref{GammaB+-fbar}) and~(\ref{GammaBfbar}), one has
\beq\label{CPK}
\langle \Gamma(B \to f_{{\rm CP}\pm} K_i) \rangle  =
2A^2_{{\rm CP}\pm} \left [ X_i \pm Y_i\cos\gamma \right ]~,
\qquad (i = c, n)~,
\eeq
where the two signs on the right-hand-side correspond to positive and negative
CP-eigenstates. Adding and subtracting rates for even-CP and odd-CP 
eigenmodes, one finds
\beqa\label{SigmaCP}
\Sigma^i_{\rm CP} & \equiv & 
\frac{\langle \Gamma(B \to f_{{\rm CP}+} K_i) \rangle}{2A^2_{{\rm CP}+}}
+ \frac{\langle \Gamma(B \to f_{{\rm CP}-} K_i) \rangle}{2A^2_{{\rm CP}-}}
= 2X_i~,\\
\label{DeltaCP}
\Delta^i_{\rm CP} & \equiv &
\frac{\langle \Gamma(B \to f_{{\rm CP}+} K_i) \rangle}{2A^2_{{\rm CP}+}}
- \frac{\langle \Gamma(B \to f_{{\rm CP}-} K_i) \rangle}{2A^2_{{\rm CP}-}}
= 2Y_i\cos\gamma~,\qquad (i = c, n)~.
\eeqa
These definitions apply in practice to sums over individual CP states. 
Eq.~(\ref{SigmaCP}) provides the most direct way to
determine $X_c$ and $X_n$.
We define CP-{\it conserving} rate asymmetries between even and odd  CP-states,
\beq
{\cal A}^i_{\rm CP} \equiv \frac{\Delta^i_{\rm CP}}{\Sigma^i_{\rm CP}}~,
\qquad (i = c, n)~.
\eeq
Using Eq.~(\ref{Y<X}), one obtains two potential inequalities for
$\cos\gamma$ in terms of these ratios,
\beq
|\cos\gamma| > |{\cal A}^i_{\rm CP}|~,\qquad (i = c, n)~.
\eeq
This inequality holds separately for charged and neutral $B$
decays. The inequality for neutral $B$ decays was noted in~\cite{Fleischer}.

As mentioned above, in order to determine $\gamma$ from untagged
neutral $B$ decays, one needs in addition to the two rates for CP
eigenstates given in~(\ref{CPK}) ($i=n$) two rate measurements for a
non-CP state $f_D$ and its CP-conjugate $\bar f_D$. These rates are
given by the two equations in~(\ref{GammaBfbar}). We denote the sum
and difference of these rates and of the corresponding rates in charged 
$B$ decays by
\beqa\label{Sigmaf}
\Sigma^i_f & \equiv & \frac{\langle \Gamma(B \to \bar f_D K_i) \rangle
+ \langle \Gamma(B \to f_D K_i)\rangle}{A^2_f(1 + r^2_f)}~,\nonumber\\
\label{Deltaf}
\Delta^i_f & \equiv & \frac{\langle \Gamma(B \to \bar f_D K_i) \rangle
- \langle \Gamma(B \to f_D K_i)\rangle}{A^2_f(1 + r^2_f)}~,\qquad (i = c, n)~,
\eeqa
which imply CP-violating asymmetries, 
\beq
{\cal A}^i_f \equiv \Delta^i_f/\Sigma^i_f~,\qquad (i = c, n)~.
\eeq 
It is then straightforward to show that
$\tan^2\gamma$ is proportional to $({\cal A}^i_f)^2$ and is given by
\beq\label{tangamma}
\tan^2\gamma = \frac{(\Delta^i_f)^2}{\rho^2_f(\Delta^i_{\rm CP})^2 -
(\Sigma^i_{\rm CP} - \Sigma^i_f)^2}
= \frac{({\cal A}^i_f)^2 (\Sigma^i_f/\Sigma^i_{\rm CP})^2}
{\rho^2_f({\cal A}^i_{\rm CP})^2 - (1 - \Sigma^i_f/\Sigma^i_{\rm CP})^2}~,
~~~~(i = c, n)~,
\eeq
where
\beq
\rho_f \equiv \frac{2r_f}{1 + r^2_f}~.
\eeq
This result applies to both charged and neutral $B$ mesons.
We stress that when determining
$\gamma$ we do not rely on separating the two terms $A^2_i$
and $A^2_i r^2_i$ contributing  to $X_i$.

While the possibility of measuring $\gamma$ from untagged $B^0$ decays
alone is interesting, the most efficient way to determine the weak phase
would be to combine information from decays of charged
$B$ decays and untagged neutral $B$ decays. 
The derivation we have just
presented applies also to charged $B$ decays alone.  One needs to measure
only the combined $B^{\pm}$ rates given in Eqs.~(\ref{GammaB+-fbar}),
without needing to separate the small $A^2_cr^2_c$
term. Thus, Eq.~(\ref{tangamma}) also gives $\tan^2\gamma$ in terms of decay
rates for combined $B^{\pm} \to D K^{\pm}$ events, where $D$ mesons
are observed in decays into an even-CP, an odd-CP eigenstates, and 
a non-CP flavorless state or a flavor state.  We see that, in principle, 
a determination of $\gamma$ does not require measuring CP asymmetries 
in $B^{\pm} \to f_{\rm CP}K^{\pm}$. These asymmetry measurements,
for even-CP and odd-CP states~\cite{DCPBelle,DCPBabar}, provide 
additional useful information. Using all these measurements together with rate measurements for untagged neutral $B$ decays will lead to a more accurate determination of $\gamma$ than when using only charged $B$ mesons.

Eq.~(\ref{tangamma}) displays an explicit dependence of $\gamma$ on
rate measurements defined in 
Eqs.~(\ref{SigmaCP}),~(\ref{DeltaCP}) and~(\ref{Deltaf}). 
We see that $\tan\gamma$ is proportional to the CP asymmetries 
${\cal A}^i_f \sim r_f\cos\delta_i\sin\delta_f\sin\gamma$ ($i = c, n$),
indicating that the sensitivity for measuring $\gamma$ increases 
with $r_f$. The sensitivity depends also on the value of 
$\delta_f$. In the extreme case that $\delta_f$ vanishes (mod $\pi$) the two rates 
in~(\ref{GammaB+-fbar}) (and in~(\ref{GammaBfbar})) become equal 
and $\gamma$ cannot be extracted. 
For the two-body flavor state $f = K^+\pi^-$, the phase $\delta_f$ 
vanishes in the SU(3) symmetry limit. However, SU(3) breaking effects 
are known to be large in $D$ decays.  Consequently, sizable values of 
$\delta_f$ have been calculated for this final state in several 
models~\cite{Refs}.   This phase can be measured at a charm 
factory~\cite{deltaf,GGR}.
There are experimental indications for 
small phases in two cases of quasi two-body states, $f = K^{*+}\pi^-$ 
where $\delta_f = (12 \pm 3)^\circ ({\rm mod}~\pi)$ was 
measured~\cite{BelleDalitz}, and $f = \rho^+\pi^-$ where 
$\delta_f = (4 \pm 3 \pm 4)^\circ$ was measured~\cite{CLEOD}.
These  results were obtained by studying the Dalitz plot 
of $D^0 \to  K_S\pi^+\pi^-$ and $D^0 \to \pi^+\pi^-\pi^0$, 
respectively. No measurement exists for $\delta_f$ in $D^0 \to
K^{*+}K^-$, which can be measured 
by studying the Dalitz plot of $D^0 \to K^+K^-\pi^0$~\cite{RS}. 

As we note in the next section, a complete Dalitz plot analysis of 
three body $D^0$ decays involves other strong phases, 
which are large in regions where two resonances overlap or when 
resonances interfere with non-resonant contributions.  These 
phases will be shown to be useful when studying $\gamma$ in 
$B \to DK$, where the $D$ meson is observed in a three-body final 
state. 

%%%%%%%%%%%%%%%%%%%%%%%%%%%%%%%%%%%
%%%%%%%%%%%%%%%%% IV  %%%%%%%%%%%%%%%%
%%%%%%%%%%%%%%%%%%%%%%%%%%%%%%%%%%%
\section{$B\to DK_S$ observed in multi-body $D$ decays}\label{multy}
The study of untagged $B^0\to DK_S$ presented in the previous section
for $D$ mesons decaying in two-body modes may be extended to multibody
decays. 
To be specific, we focus on the case of the
three-body $D$ decay, 
\beq
D  \to K_S \pi^- \pi^+,
\eeq
following the discussion of $B^{\pm}\to D K^{\pm}$ in~\cite{ggsz}. In order 
to make our point, we start with a model-independent approach. We also 
explain how modeling the amplitude for $D^0 \to K_S \pi^- \pi^+$ in
terms of a sum of a given set of intermediate resonances~\cite{BW}, as done 
recently in~\cite{BelleDalitz}, may help in 
reducing the experimental error in $\gamma$.

We denote the amplitude for $D^0 \to K_S \pi^- \pi^+$ at a given
point in the Dalitz plot by
\beq\label{Dkpipi}
A(D^0 \to K_S(p_1) \pi^-(p_2) \pi^+(p_3)) \equiv
A(s_{12},s_{13}) e^{i\delta(s_{12},s_{13})}~,
\eeq
where $s_{ij} \equiv (p_i+p_j)^2$. As in two body decays, we use the convention 
$A(s_{12},s_{13}) \ge 0$ and $0 \le \delta(s_{12},s_{13}) \le 2\pi$.  
Assuming that CP is conserved in this decay, one has
\beq\label{Dbarkpipi}
A(\DbarZ \to K_S(p_1) \pi^-(p_2) \pi^+(p_3)) = A(D^0 \to K_S(p_1)
\pi^-(p_3) \pi^+(p_2))
\equiv A(s_{13},s_{12})\,e^{i\delta(s_{13},s_{12})}~.
\eeq
That is, the (complex) decay amplitude for $\DbarZ$ at a given point
$(s_{12},s_{13})$ in the Dalitz plot equals the decay amplitude for
$D^0$ at a point $(s_{13},s_{12})$ obtained by reflection across
a symmetry axis corresponding to exchanging the momenta of 
the two pions.

The density of events in the $D$ decay Dalitz plot for untagged $B \to
(K_S\pi^-\pi^+)_D K_S$ is obtained using Eqs.~(\ref{An}),~(\ref{Dkpipi}) 
and~(\ref{Dbarkpipi}) (or from the time-dependence in Appendix A),
\beqa
\frac{d^2\Gamma}{ds_{12}ds_{13}}
( B^0/\bar B^0 & \to & [K_S(p_1) \pi^-(p_2) \pi^+(p_3)]_D K_S ) =
A^2_n \left [\left (A^2(s_{12},s_{13}) + A^2(s_{13},s_{12})\right)(1 +
r^2_n) \right.
\nonumber  \\
\label{dGamma}
& &
+\left. 4r_nA(s_{12},s_{13})A(s_{13},s_{12})\cos(\delta(s_{12},s_{23})
- \delta(s_{13},s_{12}) + \gamma)\cos\delta_n \right ]~.
\eeqa
This is in complete analogy with the first of Eqs.~(\ref{GammaBfbar}). The
second equation, describing the density at the point of reflection
across the symmetry axis, involves an opposite sign for
$\gamma$. Integrating~(\ref{dGamma}) over an area (a bin) $i$ lying
below the symmetry axis and over a corresponding symmetry-reflected
area $\bar{i}$ lying above the symmetry axis, one has\footnote{The
method outlined here applies to any multibody $D^0$ decay with the set
of equations \eqref{Gammaa} unchanged. If the $i$-th bin is in the phase
space of a final state $f_D$, then the $\bar i$-th bin is a CP
transformed bin in the phase space of the corresponding $\bar f_D$ state.}
\beqa
\langle\Gamma_i \rangle & \equiv & \int_i d\Gamma
(B^0/\bar B^0  \to [K_S \pi^-\pi^+]_D K_S) =
X_n(T_i + T_{\bar{i}}) + 2 Y_n  [c_i \cos\gamma - s_i\sin\gamma]~,
\nonumber \\
\label{Gammaa}
\langle\Gamma_{\bar{i}}\rangle & \equiv & \int_{\bar i} d\Gamma
(B^0/\bar B^0 \to [K_S \pi^-\pi^+]_D K_S) =
X_n(T_i + T_{\bar{i}}) + 2 Y_n  [c_i \cos\gamma + s_i\sin\gamma]~,
\eeqa
where we define
\beqa\label{csT}
T_i &\equiv & \int_i ds_{12}ds_{13} A^2(s_{13},s_{23})~,\nonumber \\
c_i & \equiv & \int_i ds_{12}ds_{13} A(s_{12},s_{13})A(s_{13},s_{12})
\cos(\delta(s_{12},s_{23}) - \delta(s_{13},s_{12}))~,\nonumber \\
s_i & \equiv & \int_i ds_{12}ds_{13} A(s_{12},s_{13})A(s_{13},s_{12})
\sin(\delta(s_{12},s_{23}) - \delta(s_{13},s_{12}))~.
\eeqa
The partial rates $T_i$ in $D$ decays may be measured using
flavor-tagged $D^0$ decays
and are assumed to
be known. The other $D$ decay variables, $c_i$ and $s_i$, which in
principle can be measured model-independently at a charm factory 
(up to a sign ambiguity in $s_i$),  will nonetheless be taken as
unknown. Consider $k$ different bins $i$ lying below the symmetry
axis, each contributing two unknowns $c_i$ and $s_i$.  Together with 
$X_n$, $Y_n$ and $\gamma$, there are $2k + 3$ unknowns. 
Eqs.~(\ref{Gammaa}), which provide $2k$ measurables 
($\langle\Gamma_i\rangle$ 
and $\langle\Gamma_{\bar{i}}\rangle$), are therefore unsolvable.

The situation changes when one measures another neutral $B$ decay of
this type, e.g. the sequence $B^0 \to D^* K_S$, $D^* \to D^0\pi^0$,
$D^0 \to K_S\pi^+\pi^-$, which introduces a pair of new variables
analogous to $X_n$, $Y_n$. In this case one has $4k$ measurables for
$2k + 5$ unknowns, a solution for which requires $k \ge 3$. That is,
$\gamma$ may be determined by measuring partial rates in the two
untagged neutral $B$ decay modes, $B \to DK_S$ and $B\to D^* K_S$, for
at least three pairs of Dalitz plot bins in $D\to K_S\pi^+\pi^-$.

A more powerful approach is to combine information from all the
untagged neutral $B$ decays and charged $B$ decays, using both
multibody and two-body $D$ decays. For instance, in analogy with
Eqs.~(\ref{GammaB-f}), the decays $B^{\pm} \to (K_S\pi^+\pi^-)_D
K^{\pm}$ provide four measurables for each bin~\cite{ggsz},
instead of the two in~(\ref{Gammaa}). Combining charged and untagged
neutral $B\to DK$ decays, where $D \to K_S\pi^+\pi^-$, yields $6k$
measurables for $2k + 6$ unknowns, $c_i, s_i, A_c, r_c, \delta_c, X_n,
Y_n$ and $\gamma$. Therefore, two pairs of bins provide an
overconstrained system of equations for determining $\gamma$.

The two equations~\eqref{Gammaa} are not mutually independent when 
$s_i=0$,  in analogy with the singularity noted in Eq.~(\ref{tangamma})
when $\delta_f = 0$ (mod $\pi$).
However, the relevant strong phase differences which determine $s_i$ 
are large at least in some regions of the Dalitz plot of $D^0 \to K_S\pi^+\pi^-$.
Consider, for instance, the two overlapping regions of a vertical band
describing the Cabbibbo-allowed mode $K^{*-}\pi^+$ and a horizontal band
describing the doubly Cabibbo-suppressed mode $K^{*+}\pi^-$ with a 
diagonal band representing the Cabibbo-allowed mode 
$\rho^0\bar K^0$. 
The local strong phases which determine $s_i$ for these two regions 
are the phase differences between amplitudes describing the sum of 
the $K^{*-}\pi^+$ and $\rho^0\bar K^0$ contributions and the sum of the 
 $K^{*+}\pi^-$ and $\rho^0\bar K^0$ contributions. These phases are large 
and vary a lot over the overlapping regions because of the two largely 
different $K^*\pi$ contributions in the two amplitudes. 

One can reduce the number of unknowns appearing in the determination
of $\gamma$, if the unknowns coming from the
$D$ decay, $c_i$ and $s_i$, appearing in \eqref{Gammaa}, are
determined independently. 
This can be done by assuming a Breit-Wigner (BW) form for the 
intermediate resonances contributing to this decay~\cite{ggsz}. The
parameters of the model describing the $D$ decay amplitude
can then be fitted to data of tagged $D$ decays, which are abundant 
at $B$-factories. The observables in \eqref{Gammaa} now depend 
only on three unknowns, $X_n$, $Y_n$ and $\gamma$.

It is hard to quantify the theoretical error introduced by assuming a
BW form. 
One way to proceed is to change the number of resonances and see
how the sensitivity changes. This is only a partial determination of
the error. Another source of error is the accuracy of the BW assumption. 
This can be determined by the goodness of the fit to the tagged $D$ 
decays, or by using a different model for resonances, such as  a K-matrix 
model for wide resonances~\cite{Link:2003gb}. 
A rough estimate of the theoretical error caused by 
assuming a superposition of BW amplitudes is about 
$10^\circ$~\cite{BelleDalitz}. Further studies are required in order
to evaluate possible contributions of non-BW terms in the $D$ 
decay amplitude and their effect on determining $\gamma$.
As mentioned, this model-dependence can be avoided by measuring 
the parameters $c_i$ and $s_i$ at a charm factory.

%%%%%%%%%%%%%%%%%%%%%%%%%%%%%%%%%
%%%%%%%%%%%%%  IV   %%%%%%%%%%%%%
%%%%%%%%%%%%%%%%%%%%%%%%%%%%%%%%%
\section{Using isospin and neglecting an annihilation amplitude}\label{isospin}
As already mentioned, the most powerful approach for improving the 
determination of $\gamma$ is to combine charged $B$ decays with the 
information from untagged neutral $B$ decays. 
Adding untagged neutral $B$ decays to a sample of charged $B$ decays with 
the same $D$ decay final states introduces only  two unknown parameters, 
$X_n$ and $Y_n$. Here we discuss an approximation which may be used to 
reduce the number of parameters further. This introduces a 
theoretical error in $\gamma$. Yet, it is worthwhile considering such 
an approximation as long as this error is smaller than  the statistical error.

We recall two isospin relations~\cite{isospin}, one for $\bar b \to 
\bar c u \bar s$ transitions,
\beq\label{isospin-c}
A(B^0 \to D^- K^+) = A(B^+ \to \bar D^0 K^+)- A(B^0 \to \bar D^0 K^0)~,
\eeq
and another for $\bar b\to \bar u c \bar s$ transitions,
\beq\label{isospin-u}
A(B^0 \to D^0 K^0) = A(B^+ \to D^0 K^+)+ A(B^+ \to D^+ K^0)~.
\eeq
The amplitude of $B^+ \to D^+ K^0$ is pure annihilation, and is
expected to be smaller than the other two amplitudes in the last
relation~\cite{SU3}.  The absence of rescattering effects, 
which may enhance this amplitude to a level comparable to the other two
amplitudes in this relation, can be tested~\cite{BGR} by setting very
stringent experimental bounds on the branching ratio for $B^+ \to D^+ K^0$.
Neglecting $A(B^+ \to D^+ K^0)$, Eq.~(\ref{isospin-u}) reduces to~\cite{isospin}
\beq \label{isospin-u-im}
A(B^0 \to D^0 K^0) = A(B^+ \to D^0 K^+)~.
\eeq
Namely, the two color-suppressed amplitudes have equal magnitudes and
equal strong phases. Note that the error due to isospin violation, at most a
few percent, is likely to be much smaller than that involved in neglecting the
annihilation amplitude.

Equations~(\ref{isospin-c}) and~(\ref{isospin-u-im}) may be used
to simplify the determination of $\gamma$ when combining $B^{\pm}$ decays
and untagged $B^0$ decays. The two equations imply
%MG corrected
\beqa
A^2_c + 2A^2_n - 2\sqrt{2} A_c A_n \cos(\delta_n - \delta_c) & = &
 \Gamma(B^0 \to D^-K^+)~,\nonumber \\
\label{Iso-relations}
\sqrt{2}A_n r_n & = & A_c r_c~.
\eeqa
The right-hand side of the first equation, which involves a color-allowed process,
has already been measured~\cite{Abe:2001wa} and will be assumed to be given.   
Eqs.~(\ref{Iso-relations}) reduce the six parameters
describing charged and neutral $B \to DK$ decays, $A_i, r_i, \delta_i
(i = c, n)$, to four independent ones. The measurable parameters in
untagged $B^0 \to DK_S$ decays, $X_n \equiv A_n^2(1 + r_n^2)$ and $Y_n
\equiv 2A_n^2 r_n\cos\delta_n$, can now be expressed in terms of the
three $B^+$ decay parameters and a single $B^0$ decay parameter
($\delta_n$, for instance). That is, under the above assumption,
adding information from untagged neutral $B^0$ decays to studies of
$B^{\pm}$ decays involves a single new unknown parameter instead of
two parameters. This is expected to reduce the statistical error in
determining $\gamma$.

%%%%%%%%%%%%%%%%%%%%%%%%%%%%%%%
%%%%%%%%%%%%%  V  %%%%%%%%%%%%%
%%%%%%%%%%%%%%%%%%%%%%%%%%%%%%%
\section{Concluding remarks}

Before concluding let us make several general comments:
\begin{itemize}
\item Since the data set of $B^+ \to DK^+$ is larger 
than that of $B^0 \to DK_S$, hadronic parameters such as $X_c$ 
are easier to measure than $X_n$.
As noted, the approximate relation $\sqrt{2}A_nr_n \simeq A_cr_c$
implies comparable sensitivities to $\gamma$ in charged and
neutral $B$ decays. The sensitivity in $B^0$ decays is smaller by a factor 
$\sqrt{2}$ since only $K_S$ mesons are experimentally useful. 
Moreover, the detection efficiency for $K_S$ is about a factor of two 
smaller than that for charged kaons. This is expected to reduce 
somewhat the effect of neutral $B$ decays on determining $\gamma$.  
\item Our study focused on $B \to DK$ decays. It can be extended to 
multibody $B$ decays, including $B^+\to D^*K^+$ and $B^0\to D^*K_S$,
where $D^* \to D\pi^0$, as well as to the self-tagged decays 
$B^+\to D^{(*)}K^{*+}$ and $B^0 \to D^{(*)} K^{*0}$. This would add to 
the statistical power of the analysis since the parameters related to $D$ 
decays are common to these processes and to $B \to DK$.
\item The method we discussed for $B^0$ decays can be applied also 
to $B_s$ decays, replacing the $K_S$ by $\phi, \eta', \eta$. 
In that case the advantage of being able to use untagged data is greater, 
because the hadronic environment where $B_s$ decays will be studied makes 
flavor tagging less efficient.  We have neglected the width difference
between the two neutral $B$ meson states, which is a very good 
approximation for nonstrange $B$ mesons. In the case of $B_s$, the width 
difference is expected to be nonnegligible and may be taken into account in a 
straightforward manner.
\item In our discussion we assumed that strong phases in $D$ decays  are
unknown and need to be determined from the analysis simultaneously with 
$\gamma$.  As we did already mention, the strong phases in two-body 
and quasi two-body $D$ decays can be determined 
independently~\cite{Atwood:2003mj,deltaf,GGR,RS}. 
The strong phases in three-body decays may also be determined by assuming that 
the $D$ decay amplitude is given as a sum of Breit-Wigner amplitudes.
Knowledge of strong phases would imply, for instance, that fewer $D$ 
decay modes are needed in order to determine $\gamma$ from untagged 
decays alone. In practice, this implies that a combined fit of the data to
fewer hadronic parameters will result in a smaller error in 
$\gamma$.
\item The extraction of $\gamma$ from $B^{\pm} \to DK^{\pm}$ involves 
a number of discrete ambiguities~\cite{GW,ambig}.
The ambiguities in untagged $B^0$ decays may be identified in 
Eqs. \eqref{GammaBfbar} which are invariant under
\begin {eqnarray}
\Pex &\equiv & \{\gamma\to\delta_f \,, \ \delta_f\to\gamma\}\;,  \nonumber \\
\Psign &\equiv & \{\gamma\to-\gamma\,, \ \delta_f\to-\delta_f\}\;,
\nonumber \\
\Ppi &\equiv & \{\gamma\to\gamma +\pi\,, \ \delta_f\to\delta_f +\pi
\quad {\rm or} \quad
\delta_n\to \delta_n+\pi\}\;.
\end {eqnarray}
Once several two-body $D$ decay modes are combined, or once a multibody 
$D$ decay mode is used, the first two ambiguities may be resolved.
The ambiguity $\Pex$ is lifted, since $\delta_f$ is not expected to be the 
same for all two-body and quasi two-body $D$ decay modes, and is known 
to change over the Dalitz plot in three-body decays.
Measuring the sign of $s_i$ in
\eqref{Gammaa} through a fit to a sum of Breit-Wigner resonance 
functions would resolve the $\Psign$ ambiguity.
This introduced essentially no model-dependence, since one needs only
the sign of $s_i$, which is easily determined in the vicinity of a BW resonance.
Resolving $\Psign$ avoids the ambiguity $\gamma \to \pi - \gamma$, a 
combination of $\Psign$ and $\Ppi$, which is particularly problematic 
in view of the proximity of $\gamma$ to $\pi/2$~\cite{ambig}. 
The only remaining ambiguity is $\gamma\to\gamma+\pi$. This ambiguity 
is the least problematic, since the two corresponding values of $\gamma$ 
are maximally separated.
\end{itemize}

To conclude, we have studied the information obtained from untagged neutral 
$B$ decays of the type $B \to DK_S$ involving several $D$ decay modes.
We have shown that these measurements alone can, in principle, 
determine $\gamma$. Of course, $B^0$ tagging information, while
limited, can only improve this determination. By combining information 
from untagged $B^0$ decays with that obtained in corresponding charged 
$B$ decays one gains statistics,
thereby permitting a more accurate determination of $\gamma$. 
While statistics are limited, one may neglect an annihilation amplitude in 
$B \to DK$, reducing by one the number of hadronic 
parameters and resulting in a smaller experimental error in $\gamma$.
This introduces a theoretical error in $\gamma$ 
which must be further studied.

\acknowledgments
The work of Y. G. is supported in part by the Department of Energy
under contracts DE-AC03-76SF00515 and DE-FG03-92ER40689. 
The work of A. S. is supported by the U.S. Department of Energy under 
contract DE-FG03-93ER40788. The work of J. Z. is supported in part 
by EU grant HPRN-CT-2002-00277 and  by the Ministry 
of Education, Science and Sport of the Republic of Slovenia.

\appendix
%%%%%%%%%%%%%%%%%%%%%%%%%%%%%%%
%%%%%%%%%%%%%  A  %%%%%%%%%%%%%
%%%%%%%%%%%%%%%%%%%%%%%%%%%%%%%
\section{Time-dependent $B \to DK_S$ with multibody $D$ decays}
%%%%%%%%%%%%%%%%%%%%%%
\def\Dmbt{{\Delta m_{\sss B} \, t \over 2}}
\def\d{\displaystyle}
\def\sss{\scriptscriptstyle}
\def\B0bar{\Bbar{}^0}
\newcommand{\Bbar}{\,\overline{\!B}}
\def\bear{\begin{array}}
\def\enar{\end{array}}
In this appendix we provide a formalism allowing the extraction
of $\gamma$ from time-dependent rates in $B^0\to f_D K_S$
where $f_D$ is a multibody final state. As we show, this
does not only serve the purpose of determining $\gamma$, but
also helps resolve the current two-fold ambiguity, 
$\beta \to \pi/2 - \beta$.
For simplicity we take $f_D$ to be the three-body final
state $K_S \pi^+\pi^-$ studied in Section IV.

Time-dependent partial rates, integrated over a bin $i$ in the Dalitz
plot of $D \to K_S\pi^+\pi^-$ lying below the symmetry axis, and over 
a corresponding symmetry-reflected bin $\bar i$ above the axis, 
are readily calculated for initial $B^0$ and $\bar B^0$ states
(a positive $B$ meson bag parameter is 
assumed~\cite{Grossman:1997xn}):
\beqa
 \Gamma_i &\equiv & \int_i d \Gamma \left(B^0 (t) \to (K_s \pi^-\pi^+)_D
 K_s \right) = e^{- \Gamma_{\sss B} t} \,
 A_n^2 \times \nonumber \\
&~& \left\{ I^+_i \cos^2 \left( \Dmbt \right)+ I^-_{\bar i}
\sin^2\left(\Dmbt\right) +S_i \sin(\Delta m_{\sss B} t)
\right\}~, \label{eq:mb1}\\
\bar{\Gamma}_i &\equiv & \int_i d \Gamma \left(\bar B^0 (t) \to (K_s
\pi^-\pi^+)_D K_s \right) = e^{- \Gamma_{\sss B} t} \,
A_n^2 \times \nonumber \\
&~& \left\{ I^-_{\bar i} \cos^2 \left( \Dmbt \right)+ I^+_i
\sin^2\left(\Dmbt\right) - S_i \sin(\Delta m_{\sss B} t)
\right\}~,\\
 \Gamma_{\bar i} &\equiv & \int_{\bar i} d \Gamma \left(B^0 (t) \to (K_s \pi^-\pi^+)_D
 K_s \right) = e^{- \Gamma_{\sss B} t} \,
 A_n^2 \times \nonumber \\
&~& \left\{ I^+_{\bar i} \cos^2 \left( \Dmbt \right)+ I^-_i
\sin^2\left(\Dmbt\right) + S_{\bar i} \sin(\Delta m_{\sss B} t)
\right\}~,\\
 \bar{\Gamma}_{\bar i} &\equiv & \int_{\bar i} d \Gamma \left(\bar 
B^0 (t) \to (K_s \pi^-\pi^+)_D
 K_s \right) = e^{- \Gamma_{\sss B} t} \,
 A_n^2 \times \nonumber \\
&~& \left\{ I^-_i \cos^2 \left( \Dmbt \right)+ I^+_{\bar i}
\sin^2\left(\Dmbt\right) - S_{\bar i} \sin(\Delta m_{\sss B} t)
\right\}~. \label{eq:mb4} 
\eeqa
The six observables determined from the time-dependence,
$I^{\pm}_{i}, I^{\pm}_{\bar i}$, $S_{i}$ and $S_{\bar i}$, are defined in terms of 
the quantities in Eqs.~(\ref{csT}),
 \beqa \label{observablesI}
I^{\pm}_i &\equiv & T_{\bar i} + r^2_n T_i + 2 r_n
\left[\cos(\gamma \pm \delta_n)c_i \mp \sin(\gamma \pm
\delta_n)s_i \right]~, \\ \label{observablesS}
S_i &\equiv & r_n T_{\bar i}\sin(2\beta+\gamma -
\delta_n) + [\sin(2\beta)c_i - \cos(2\beta)s_i] \nonumber \\ &+&
r^2_n[\sin(2\beta +2\gamma)c_i + \cos(2\beta+2\gamma)s_i] + r_n T_i
\sin(2\beta+\gamma + \delta_n)~.
\eeqa 
Expressions for the observables $ I^{\pm}_{\bar i}$ and $S_{\bar i}$ 
are obtained from (\ref{observablesI}) and (\ref{observablesS})
by replacing $T_i\leftrightarrow T_{\bar{i}},~s_i \to -s_ i$. Note that
$I^\pm_{i,\bar i}$ correspond directly to the partial decay widths
$\Gamma^\pm_{i,\bar i}$ for $B^\pm\to D K^\pm$ defined
in~\cite{ggsz}.

Dividing the Dalitz plot into $k$ pairs of bins, $i$ and $\bar i$,
the $6k$ observables permit an extraction of $\gamma$. There are 
$2k + 4$ unknowns, $ c_i, s_i, A_n, r_n, \delta_n, \gamma $ 
($\sin(2\beta)$ is assumed to be known), so that the system is 
solvable for $k \geq 1$. Namely, in order to determine $\gamma$ 
from time-dependent decay rates into $(K_S\pi^+\pi^-)_DK_S$,
it is sufficient to divide the $D$ decay Dalitz plot into two bins, symmetric 
with respect to the symmetry axis.

The solution for $\gamma$ involves a four-fold discrete ambiguity.
Equations~(\ref{eq:mb1})--(\ref{eq:mb4}) are invariant under the following
four independent discrete transformations:
\beqa
P^{\gamma}_{\pi} &\equiv& \{\gamma \to \gamma + \pi, \delta_n \to
\delta_n + \pi\}~, \nonumber \\
P'_{\pi} &\equiv& \{\gamma \to \gamma + \pi, \beta
\to \beta + \pi/2, c_i \to -c_i, s_i \to -s_i\}~, \nonumber \\
P^{\beta}_{\pi} &\equiv&\{\beta \to \beta + \pi\}~, \nonumber \\
P_{-} &\equiv& \{\gamma
\to -\gamma, \beta \to \pi/2 - \beta, \delta_n \to - \delta_n, s_i \to
-s_i\}~.
\eeqa
The $P'_{\pi}$ ambiguity can be resolved model-independently, either 
by using the sign of $\sin 2 \beta$ or by measuring the sign of $c_i$ at a 
$\Psi (3770)$ charm factory~\cite{ggsz}.  The $P_-$
ambiguity can be resolved if one determines the sign of $s_i$ by
fitting the Dalitz plot to a sum of Breit-Wigner forms. (See Section  VI.)  
Note that resolving the $P_-$ ambiguity in this way leads to the determination
of the sign of $\cos(2 \beta)$ in an essentially model-independent way.
It also determines the sign of $\gamma$ or equivalently
the sign of $\cos(2 \alpha)$.  Fixing the sign of $\cos (2\beta)$ is 
a consequence of knowing the sign of $s_i$ in multibody decays.
This is impossible in two-body $D$ decays, where the sign of $\sin\delta_f$ 
cannot be determined. The remaining two ambiguities, $P^{\gamma}_{\pi}$ 
and $P^{\beta}_{\pi}$, cannot be resolved without further theoretical input.

%%%%%%%%%%%%%%%%%%%%%%%%%%%%%%%%%
%%%%%%%%%%%%%  bib  %%%%%%%%%%%%%
%%%%%%%%%%%%%%%%%%%%%%%%%%%%%%%%%

\end{document}